\documentclass[twocolumn,aps,reprint,]{revtex4-2}
\usepackage[T1]{fontenc}
\usepackage{epsfig,amsmath}
\usepackage{amssymb}
\usepackage{subfigure}
\usepackage{graphicx}
\usepackage[colorlinks,linkcolor=blue]{hyperref}
\usepackage{url}
\usepackage{bm}
\usepackage{mathrsfs}
\usepackage{dcolumn}
\usepackage{booktabs} 
\usepackage{multirow}
\usepackage{stmaryrd}
\usepackage{mathrsfs}
\usepackage{pifont}
\usepackage{amsthm}
\usepackage{latexsym}

\usepackage{babel}
\usepackage{lineno} 

\begin{document}
\title{Modularized and Scalable Compilation for Double Quantum Dot Quatum Computing}

\author{Run-Hong He$^{1}$, Xu-Sheng Xu$^{2*}$\footnote{thuxuxs@163.com}, Mark S. Byrd$^{3}$ and Zhao-Ming Wang$^{1\dagger}$\footnote{wangzhaoming@ouc.edu.cn}}
\thanks{$^*$thuxuxs@163.com \\ $^\dagger$wangzhaoming@ouc.edu.cn}

\address{$1.$College of Physics and Optoelectronic Engineering, Ocean University
of China, Qingdao 266100, China
\\
$2.$Department of Physics, State Key Laboratory of Low-Dimensional Quantum Physics, Tsinghua University, Beijing 100084, China
\\
$3.$Department of Physics, Southern Illinois University, Carbondale, Illinois 62901-4401, USA}

\begin{abstract}
Any quantum program on a realistic quantum device must be compiled into an executable form while taking into account the underlying hardware constraints. Stringent restrictions on architecture and control imposed by physical platforms make this very challenging. In this paper, based on the quantum variational algorithm, we propose a novel scheme to train an Ansatz circuit and realize high-fidelity compilation of a set of universal quantum gates for singlet-triplet qubits in semiconductor double quantum dots, a fairly heavily constrained system. Furthermore, we propose a scalable architecture for a modular implementation of quantum programs in this constrained systems and validate its performance with two representative demonstrations, Grover's algorithm for the database searching (static compilation) and a variant of variational quantum eigensolver for the Max-Cut optimization (dynamic compilation). Our methods are potentially applicable to a wide range of physical devices.  This work constitutes an important stepping-stone for exploiting the potential for advanced and complicated quantum algorithms on near-term devices.  
\end{abstract}
\maketitle

\section{INTRODUCTION}

Using superposition and entanglement, quantum computers could admit superpolynomial or even exponential speedup over their classical counterparts when solving certain important and otherwise intractable problems \cite{qc_nielsen,qc_bb}. In the race to construct quantum computing prototypes, a wide range of physical models have been proposed and experimentally demonstrated over the
past decades, including trapped ions \cite{trapped_ions_1}, ultracold atoms \cite{ultracold_atoms_1}, nitrogen-vacancy centers \cite{nv_center_1}, superconducting circuits \cite{zuchongzhi21},
optical system \cite{jiuzhang2} and semiconductor quantum dots \cite{qd_review_0,qd_review_1}, etc.  In the past, using electron spins in solid-state systems, semiconductor quantum dots have been favored due to the prospective scalability and compatibility with existing mature semiconductor manufacturing techniques  \cite{qd_fabrication,qd_2_ms_coherence_time_silicon,qd_6_spin_silicon_Nature,qd_3_qubit_error_correction}.
By leveraging the degrees of spin and charge of electrons trapped in quantum dots, various qubit modalities have been suggested and realized in experiments, such as the single-electron spin-1/2 qubit \cite{qd_6_spin_silicon_Nature}, two-electron singlet-triplet ($S$-$T_{0}$) qubit in double quantum dots (DQDs) \cite{qd_ST0_one_qubit_tunable_x_rate_APS}, three-electron exchange-only qubit \cite{qd_exchange_only_0} and hybrid qubit \cite{qd_hybrid_0}. Among them, the $S$-$T_{0}$ qubit has attracted considerable interest because it can be manipulated all electrically and provides rapid gates in sub-nanosecond time scales, which is fast enough to enable $10^{4}$ gates before the decoherence takes over \cite{qd_ST0_one_qubit_tunable_x_rate_APS}.

Any high-level algorithm for quantum computation needs to be translated into low-level instructions that can be executed step by step on specific quantum hardware  \cite{compiler_Nature_Review,he_sn_compiling}. While the $S$-$T_{0}$ qubit is one of the most promising modalities for constructing a scalable quantum computer from the fabrication perspective, the tight restrictions on the control limits the applications of many existing optimization methods and make it challenging to manipulate this quantum system for computational tasks 
\cite{dqn_state_preparation_npj_qi,qd_ST0_wangxin_PRA_2014}. The primary hurdles arise in physical details that \textbf{1)} one could only precisely control the rotation rate around the $+z$-axis of a single-qubit state on the Bloch sphere, while the rotation rate around the $x$-axis is constant and non-zero; \textbf{2)} the undesired coupling introduced by simultaneous operations on adjacent qubits would alter these intended individual operations. To date, an enormous amount of ingenuity and effort has been devoted to compiling required operations into sequential native gates on this platform, such as the so-called SUPCODE \cite{qd_ST0_wangxin_Nature_comm_2012,qd_ST0_wangxin_PRL_2013,qd_ST0_wangxin_PRA_2014} for generating robust quantum gates and fast geometric gates \cite{qd_ST0_fast_geometric_gate_PRA} for cancelling out the accompanied dynamical phase during evolution. In this paper a native gate refers to an operation that can be readily implemented in DQDs with a single (for single-qubit operation) or pair (for two-qubit operation) of pulses. Typically, to perform arbitrary rotation gates, it is necessary to iteratively solve a set of nonlinear equations to determine the appropriate composite pulse sequence \cite{qd_ST0_wangxin_PRA_2014,qd_ST0_supervised_learning}. This is a resource-expensive and time-demanding task in practice. A significant surge of interest has recently been focused on nascent tools that are aimed at improving the efficiency of tailoring pulse sequences.  For examples,
by employing the machine learning \cite{deep_learning_book}, Ref. \cite{qd_ST0_supervised_learning} studies how to directly predict the required pulse sequence with a well-trained neural network; Ref. \cite{dqn_state_preparation_npj_qi} promises to dynamically steer a specific quantum state to another; Refs. \cite{he_DRL,gate_decompose_prl_dqn}
could reset an arbitrary quantum state to a target one, and Ref. \cite{ppo_state_preparation} shows how to prepare an arbitrary quantum state from $|0\rangle$. Furthermore, high fidelity
universal quantum state preparation is also observed in \cite{he_RG} with the revised greedy algorithm. All of these exhibit excellent performance for universal gate sets or few-qubit state preparation. Whereas the lack of an architecture that can properly combine these pulse sequences of varying lengths makes it quite difficult to perform a practical large-scale computation due to the aforementioned second limitation in DQDs - an unfinished operation on one qubit will interfere with the execution of quantum gates on its neighboring qubits, thus resulting in the
absence of scalability. An open problem and important direction of investigation for this special system is therefore to develop an architecture which could layout various native gates such that the system can scalably compile impactful computational tasks. Available traditional optimization approaches that could yield a fixed-length pulse sequence are mainly based on the gradient, such as the stochastic gradient descent \cite{SGD}. Partly due to the fact that each gradient evaluation requires two forward passes for loss function calculation, their resource consumption grows
very quickly as the size of parameter space increases. In addition, their performance is sensitive to the initialization, and as a consequence, the optimization routine may often be stucked in a local optimum and then end with an inadequate fidelity.

In this paper, we enlist another emerging and powerful technique to compile a desired unitary into a native gate sequence without changed length - the variational quantum algorithm (VQA) \cite{tfq_white_book}, which has recently been used for decomposing complex unitary transformations into ordered universal gates \cite{gate_decompose_with_VQA,gate_decompose_with_VQA_adaptive_circuit_compression} or Schmidt decomposition \cite{Schmidt_decompose_with_VQA}, etc.
Utilizing the adjoint differentiation \cite{tfq_white_book}, the VQA permits gradients collection of all parameters with only one forward pass and one recursive backward pass \cite{tfq_white_book} - a calculation saving compared to traditional gradient-based methods. Additionally, we suggest training the Ansatz, a circuit of parametric native gates, with random quantum states. Compared to the widely used matrix-based approaches \cite{gate_decompose_with_VQA,gate_decompose_with_VQA_adaptive_circuit_compression,Schmidt_decompose_with_VQA}, this training scheme offers reduced overhead in loss calculation and smaller possibility of being stucked in local optimums during optimization. Considering single-qubit rotations and entangling two-qubit gates are crucial ingredients for universal quantum computation, a series of them are compiled and high fidelities are reached, underling the foundations for accurately implementing quantum programs in DQDs.

More importantly, inspired by the special stacking style of Mahjong cards, a traditional Chinese game (see Fig.~\ref{fig:2}(a)), we present a modularized and scalable (MS) compilation architecture that can systematically combine native gate sequences into logic operations to perform advanced and complex reference quantum circuits in DQDs. Finally, we demonstrate our MS compilation with two representative quantum computing tasks, i.e., the Grover's algorithm \cite{grover_0,grover_3} for database search and a variant \cite{MBE_for_max_cut} of variational quantum eigensolver (VQE) \cite{VQE_photon,VQE_big_size_Nature_Comm_2022} for graph Max-Cut optimization \cite{max_cut}, and achieve excellent results. These results show that our MS compilation can be a powerful tool used to exploit the potential of DQDs in the current noisy intermediate-scale quantum devices era \cite{NISQ}. \\

\section{Model\label{sec:Model}}

In the present work, the model of interest is the semiconductor DQDs system, where the $S$-$T_{0}$ qubit is encoded in the collective spin states of two electrons confined in a double-well potential  \cite{qd_ST0_one_qubit_tunable_x_rate_APS,qd_ST0_three_qubits}. The Hamiltonian of a single $S$-$T_{0}$ qubit, governed by external electric pulses is \cite{qd_ST0_one_qubit_micromagent_PNAS_2014,qd_ST0_one_qubit_Science,qd_review_0,qd_review_1}
\begin{equation}
H(t)=J(t)\sigma_{z}+h\sigma_{x},\label{eq:0}
\end{equation}
in the computational basis $\{|0\rangle\!\equiv\!|S\rangle\!=\!(|\uparrow\downarrow\rangle-|\downarrow\uparrow\rangle)/\sqrt{2}$, $|1\rangle\!\equiv\!|T_{0}\rangle\!=\!(|\uparrow\downarrow\rangle+|\downarrow\uparrow\rangle)/\sqrt{2}\}$.
$\sigma_{z}$ and $\sigma_{x}$ are Pauli matrices and represent rotations of the quantum state around the $z$- and $x$-axes with rates $J(t)$ and $h$, respectively. The coefficient $h$ describes 
the Zeeman energy splitting between $|S\rangle$ and $|T_{0}\rangle$, commonly raised by nearby deposited permanent micromagnet \cite{qd_ST0_one_qubit_micromagent_PNAS_2014}
and its value is difficult to vary during the quantum gate time in experiment \cite{qd_ST0_one_qubit_micromagent_PNAS_2014,qd_ST0_three_qubits} (although tunable splitting by Overhauser field has been observed \cite{qd_ST0_one_qubit_tunable_x_rate_APS}, its time consumption is much longer than a typical gate time). We assume $h=1$ here to facilitate our theoretical treatment, and take this as the energy unit. The only effective tunable parameter in this system is the exchange coupling $J(t)$ between two captured electrons and can be rapidly manipulated by applying calibrated voltage pulses to the associated electrodes. In addition, for simplicity, we let the reduced Planck constant $\hbar=1$ which determines the time-scale throughout. Because of the
nature of the exchange interaction and to avoid altering the charge configuration of the quantum dots, $J(t)$ is constrained to be non-negative and bounded, i.e., $J_{\mathrm{max}}\geq J(t)\geq0$, where the maximal value $J_{\mathrm{max}}\sim h$ \cite{qd_ST0_wangxin_Nature_comm_2012}. In other words, we have only precise control over the rotation rate of the quantum state around the $+z$-axis on Bloch sphere, while the rotation rate around the $x$-axis is constant and always present.

An experimentally available native gate (one-piece rotation) can be implemented by applying an electrical pulse with specific intensity $J$ for a certain amount of time $\triangle t$, which produces a rotation around the axis $Jz+x$:
\begin{equation}
g(J,\triangle t)=\mathrm{exp}(-i(J\sigma_{z}+\sigma_{x})\triangle t).\label{eq:1}
\end{equation}
As a special case, if we set $J=0$, then $g(0,\triangle t)=\mathrm{cos}(\triangle t)\cdot I+i\mathrm{sin}(\triangle t)\cdot\sigma_{x}$, where $I$ refers to the identity operator. In this case, the identity gate $g_{I}$ can be achieved, up to an irrelevant global phase, when $\triangle t=k\pi$, where $k$ are integers. This point will prove valuable and plays the role of a major groundstone for our MS compiling architecture in DQDs as we will discuss in the following section.

In order to generate entanglement between qubits, multi-qubit entangling gates are necessary. In DQDs, the effective Hamiltonian of two adjacent qubits based on Coulomb interaction can be written as \cite{qd_ST0_2_qubit_Science_2012,qd_ST0_2_qubit_npj_qi_2017,qd_review_1}
\begin{equation}
\begin{aligned}
H_{\mathrm{2\!-\!qubit}}\!=&\frac{\hbar}{2}(J_1(\sigma_z\otimes I)\!+\!J_2(I\otimes\sigma_z)\!+\!h_1(\sigma_x\otimes I)\\ & +\! h_2(I\otimes\sigma_x)\!+\!\frac{J_{12}}{2}((\sigma_z\!-\!I)\otimes(\sigma_z\!-\!I))),\label{eq:2}
\end{aligned}
\end{equation}
in the basis of $\{|SS\rangle,|ST_{0}\rangle,|T_{0}S\rangle,|T_{0}T_{0}\rangle\}$. $J_{i}$ and $h_{i}$ are exchange interaction and Zeeman splitting respectively, with the subscript $i=1,2$ referring to the corresponding qubit.  Empirically, the inter-qubit coupling $J_{12}\propto J_{1}J_{2}$ \cite{qd_ST0_2_qubit_Science_2012} and here we assume $J_{12}=J_{1}J_{2}/2$ again for simplicity.  As in the case of single $S$-$T_{0}$ qubit, the operation on this two-qubit system requires only pulsed electric fields, i.e., $J_{1}$ and $J_{2}$. To maintain the coupling between two qubits, $J_{i}>0$ during entangling operations. It is important to notice that the coupling term in Eq.~(\ref{eq:2}) also indicates that individual qubit manipulation of a given qubit requires the suspension of the control on its neighboring qubits to avoid undesired inter-qubit coupling. These restrictions should be kept in mind for the effective design of the architecture to run quantum algorithms.

\section{Methods \label{sec:Method}}

\begin{figure*}[!htbp]
	\centering	
	\subfigure[ ]{\includegraphics[scale=0.5]{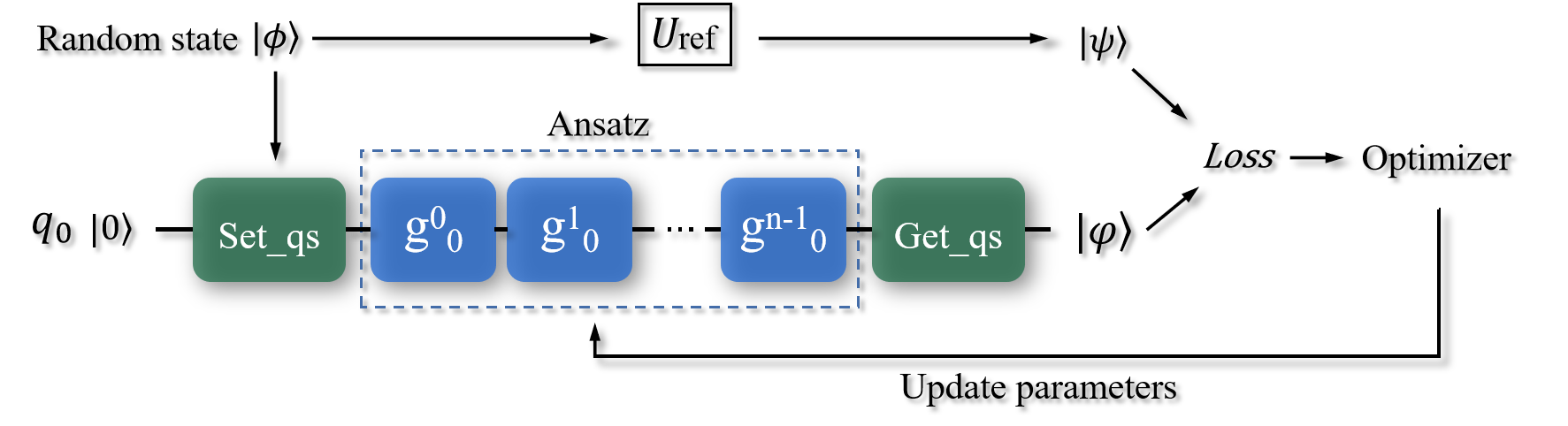}}
	\subfigure[ ]{\includegraphics[scale=0.5]{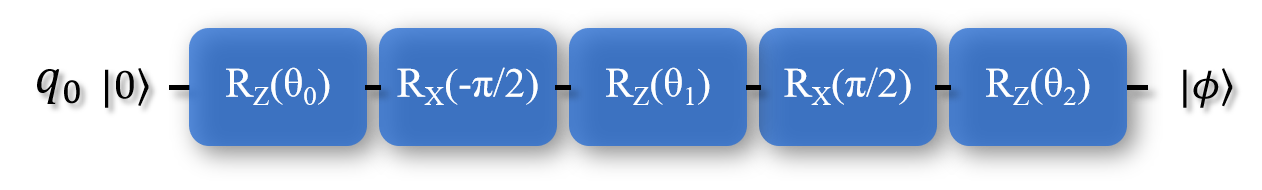}}
	\subfigure[ ]{\includegraphics[scale=0.5]{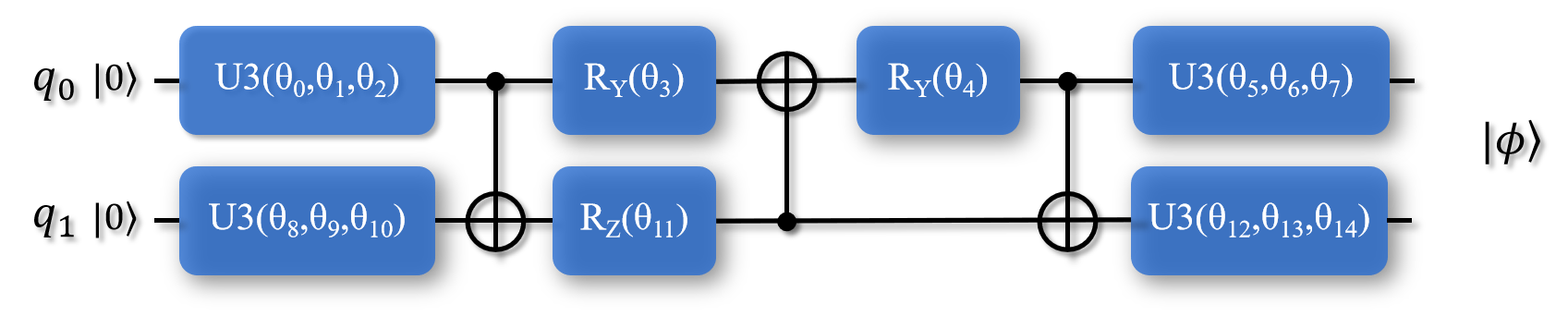}}
	\caption{(a) The overall workflow of the compilation
		of a reference logical single-qubit unitary $U_{\mathrm{ref}}$ with
		an ansatz consisting of sequential native gates. Each native gate,
		denoted as a blue block, is carried out by a single electric pulse
		$J$ with a certain duration acting on the DQD qubit. The details
		of this workflow are described in text. (b) The so-called U3 circuit
		\cite{U3_and_virtual_Z,MindQuantum} used to generate single-qubit
		random quantum states for training one-qubit ansatz. (c) The parametric
		circuit \cite{minimum_universal_two_qubit_circuit} used to generate
		two-qubit random quantum states for training two-qubit ansatz.}\label{fig:The-overall-workflow}
\end{figure*}

In the previous section, we presented a typical restricted system, $S$-$T_{0}$ qubits in DQDs and pointed out where the difficulty of gate compiling lies. In this section, we start by explaining how to compile a single-qubit arbitrary gate into sequential DQDs native gates by VQA and next describe an architecture that orchestrates operations on neighbor qubits to avoid unexpected interaction effects. Finally, a structure that enables two-qubit entangling gates is presented. With these elements, advanced quantum algorithms can be implemented on a system consisting of DQDs.

In our work, any reference unitary $U_{\mathrm{ref}}$ is compiled into DQDs native gates by leveraging the VQA. Taking the complication of an arbitrary single-qubit gate as an example, the overall workflow is schematically outlined in Fig.~\ref{fig:The-overall-workflow}(a), where the ansatz is composed of $n$ native gates, denoted as blue blocks with subscripts indicating which qubit is affected. Each native gate is generated by an electric pulse with a certain strength $J$ and duration. The process of achieving an appropriate native gates sequence that is equivalent to the $U_{\mathrm{ref}}$ goes as follows:

\textbf{Step 1}: Ansatz's variational parameters ($J$s) are all ones initialized before optimization and additionally restricted to the non-negative domain throughout the training process, accounting for the realistic physical limitation on non-negative pulse strengths as introduced in the previous sections.

\textbf{Step 2}: Certain number of random quantum states is generated using quantum circuit containing random parameters, and divided into training and validation sets. The single- and two-qubit quantum circuit generating random states in this work are shown in Fig.~\ref{fig:The-overall-workflow}(b) and (c), whose outputs can theoretically cover the entire Hilbert space as the real-valued parameters vary \cite{U3_and_virtual_Z,minimum_universal_two_qubit_circuit}.

\textbf{Step 3}: A random quantum state $|\phi\rangle$ sampled from the training set will: \textbf{1)} be fed to the ansatz which then outputs the final state $|\varphi\rangle$ after evolution; and \textbf{2)} be acted by the $U_{\mathrm{ref}}$ and converted to the target state $|\psi\rangle$. 

\textbf{Step 4}: Compute the loss function $Loss=-|\langle\psi|\varphi\rangle|^{2}$, which quantifies the fitness of the ansatz to the $U_{\mathrm{ref}}$. With the $Loss$ available, the gradient with respect to each ansatz's parameter can be calculated  \cite{tfq_white_book} and then used to update the parameters for improving the performance by an optimizer, such as the Adam \cite{adam,xieyangyang_adam}. 

\textbf{Step 5}: Validate all random states in the validation set using a process similar to the step 3 and then take the maximal $1+Loss$ among them as the current error $\epsilon$ of the ansatz relative to the reference unitary at this training step.

Repeat the sequential steps 3-5 until the error $\epsilon$ over the validation set smaller than an acceptability threshold, such as $10^{-5}$, where the effect of the ansatz will be approximately the target $U_{\mathrm{ref}}$.

Compared to schemes such as  \cite{gate_decompose_with_VQA,gate_decompose_with_VQA_adaptive_circuit_compression,Schmidt_decompose_with_VQA} where ansatz's parameters are updated in a closed-loop style, the application of random states reduces the susceptibility of the optimization to local optima akin to the usage of random samples in the training landscape of classical neural networks \cite{he_DRL}. In addition, this utilization of quantum states is more resource efficient than commonly used approaches based on matrices in loss calculation, such
as the Hilbert-Schmidt distance or other customized metrics between the associated unitaries  \cite{gate_decompose_with_VQA,gate_decompose_with_VQA_adaptive_circuit_compression,Schmidt_decompose_with_VQA}. It is worth noting that this training approach is scalable and applicable
generally to the construction or decomposition of unitary transformations with any number of qubits.  

\begin{figure*}[ht]
	\subfigure[ ]{\includegraphics[scale=0.2]{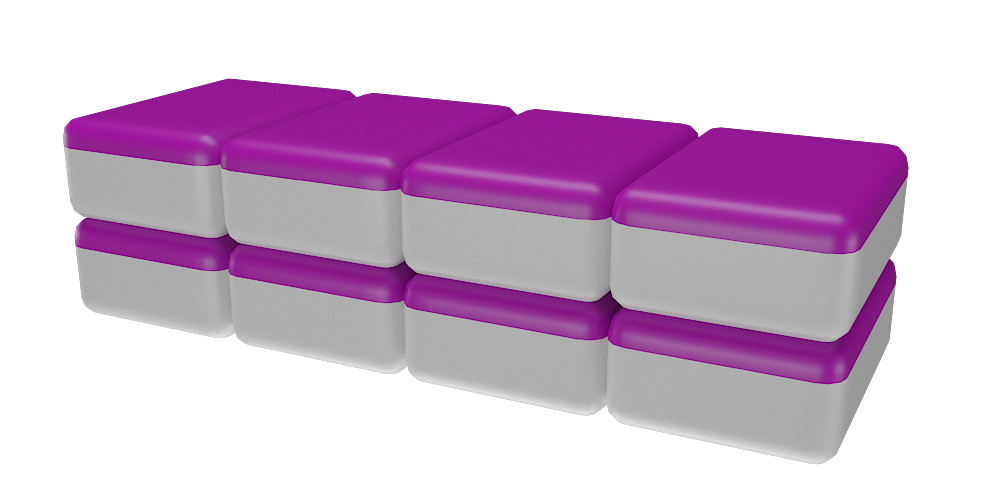}}
	\subfigure[ ]{\includegraphics[scale=0.4]{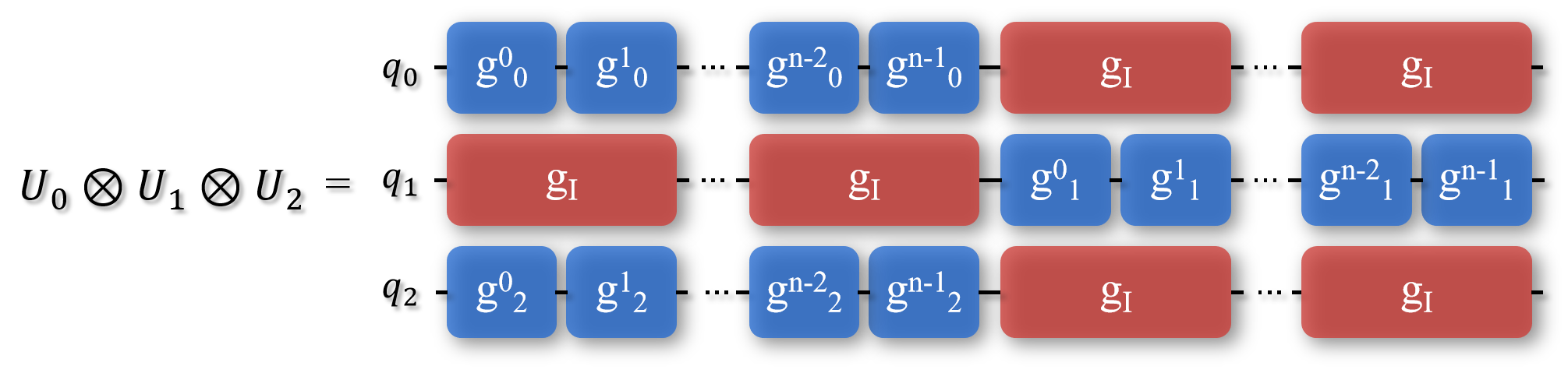}}
	\subfigure[ ]{\includegraphics[scale=0.5]{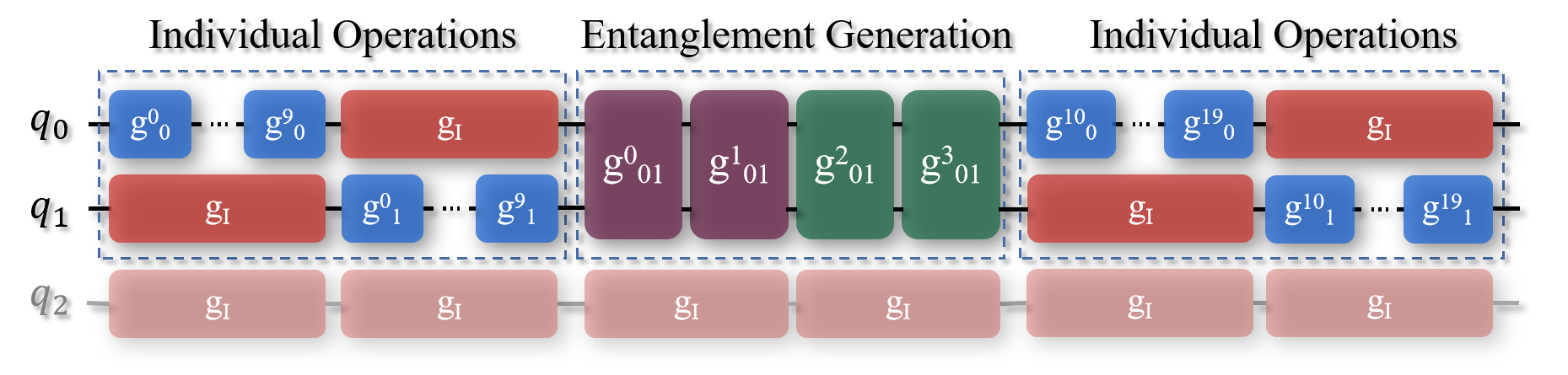}}\caption{\label{fig:2} 
		(a) The vertically stacked mahjong cards which inspired	us to explore the MS compilation. (b) Schematic diagram of the compiling architecture formed according to the MS compilation that enables interference-free control between qubits. (c) An ansatz constructed in accordance with the MS compilation which enables two-qubit entangling gates acting on $q_{0}$ and $q_{1}$, whose adjacent qubits, e.g., the $q_{2}$, remain idle during this phase to avoid undesirable interactions. The ansatz contains three parts: two Individual Operations and one Entanglement Generation, which consist of only single-parameter native gates. }
\end{figure*}

As for the implementation of individual operations on different qubits, the challenge arises from the fact that if the pulses imposed on adjacent qubits are non-zero simultaneously, these qubits will interact with each other due to capacitive coupling \cite{qd_review_1}, and therefore interrupts intended operations. Inspired by the Mahjong (shown in Fig.~\ref{fig:2}(a)) whose cards are stacked upright so that no interaction is produced between adjacent columns and so can be operated independently, we propose the following architecture to circumvent this problem.

We make the execution time of all ansatzes for the associated logical gates to be the same (constant pulse sequence length), and when a certain qubit is operated on, its neighboring qubits are set to be idle (no pulses acting on), so as to avoid interference between qubits. While the idle qubits undergo free evolution all the time due to invariable and non-zero rotation rate $h$ around the $x$-axis, as long as the gate time is chosen appropriately, these idle qubits will evolve exactly to their original status when the operations on the qubit of interest
is completed. In other words, this period of free evolution is equivalent to identity gates acting on the idle qubits. For any two adjacent qubits, there always exists one qubit with no pulse acting on it at any time, leaving them be free from the undesired coupling. With this principle, operations on adjacent qubits will be made alternately. Fig.~\ref{fig:2}(b) outlines an example which allows individual arbitrary rotation operations $U_{0}\otimes U_{1}\otimes U_{2}$ on three adjacent qubits, subscript $i=0,1,2$ referring to the corresponding qubit that is transformed.
The red blocks ``$g_{I}$'' refer to identity gates, due to the zero-strength pulses with $k\pi$ duration, where $k$ is an integer (as explained in Eq.~(\ref{eq:1})). The duration of $g_{I}$ here
is taken to be twice the typical pulse time (the duration of a parametric native gate, or blue block) for clarity.

To create two-qubit entangling gates in DQDs, e.g., CX and CZ, we likewise empirically design an ansatz form schematically visualized in Fig.~\ref{fig:2}(c), where the entangling gate acts on $q_{0}$ and $q_{1}$, while other adjacent qubits, e.g., the $q_{2}$, are left idle to avoid interference. This entanglement gate consists of three parts: one Entanglement Generation sandwiched between two Individual Operations. The part of Entanglement Generation consists of 4 sequential native 2-qubit gates (cross-qubit purple and green blocks) for generating enough entanglement between two target qubits. Considering the fact that a two-qubit native gate is brought about by two simultaneous pulses, which impose a challenge in the optimization using VQA, we fix one of them to be 1 and leave the other variable. For the first two two-qubit native gates (purple blocks) we fix their parameter on $q_{0}$, while the last two (green blocks) on $q_{1}$. The two Individual Operations parts located before and after the Entanglement Generation are used to apply additional rotations to each qubit to achieve fine tuning for an overall improved performance.  {\it This architecture is extremely important as it allows programs implemented on this platform to not only precisely execute quantum gates but also reliably implement quantum algorithms.}\\

\section{numerical Results\label{sec:Results}}

\subsection{Compilation for universal gates \label{subsec:MS-compilation-for}}

A set of universal quantum gates is necessary for the implementation of quantum programs \cite{qc_nielsen,qc_bb}.  In this subsection, we first study the applicability of the MS compilation for arbitrary single-qubit gates and then move our attention to enacting common two-qubit entangling quantum gates. The details of the gate compiling process are shown in the Section \ref{sec:Method}. We stress that, for conciseness, the control pulses are assumed to be piecewise-constant and so are turned on and off instantaneously.
In real experiments, pulses possessing finite rise times will result in only a minimally alteration in the pulse parameters yielded by our work but will not alter the conclusions as clearly explained in Ref.~\cite{qd_ST0_wangxin_Nature_comm_2012}.   

\begin{figure*}[htb]
	\subfigure[ ]{\includegraphics[scale=0.3]{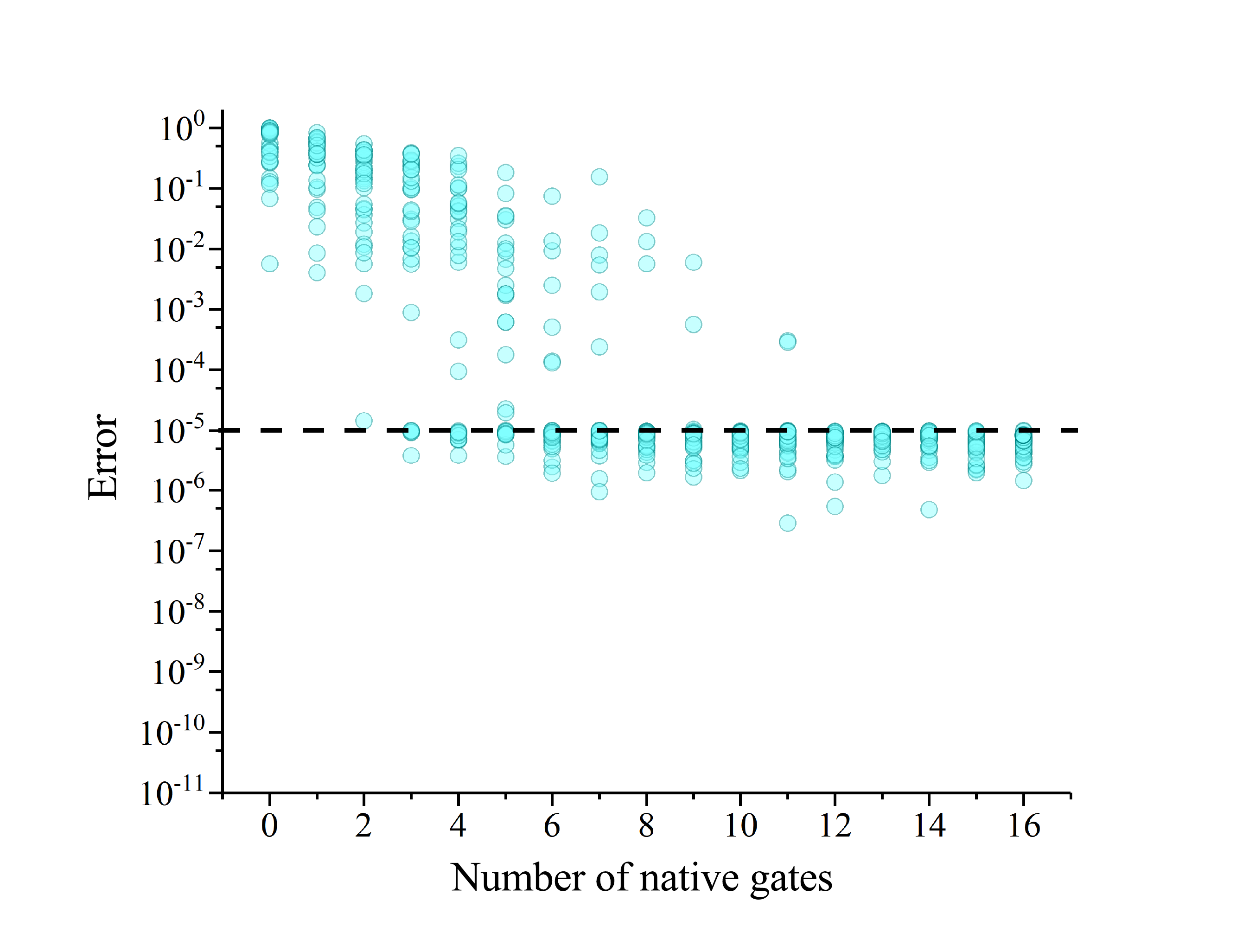}}
	\subfigure[ ]{\includegraphics[scale=0.3]{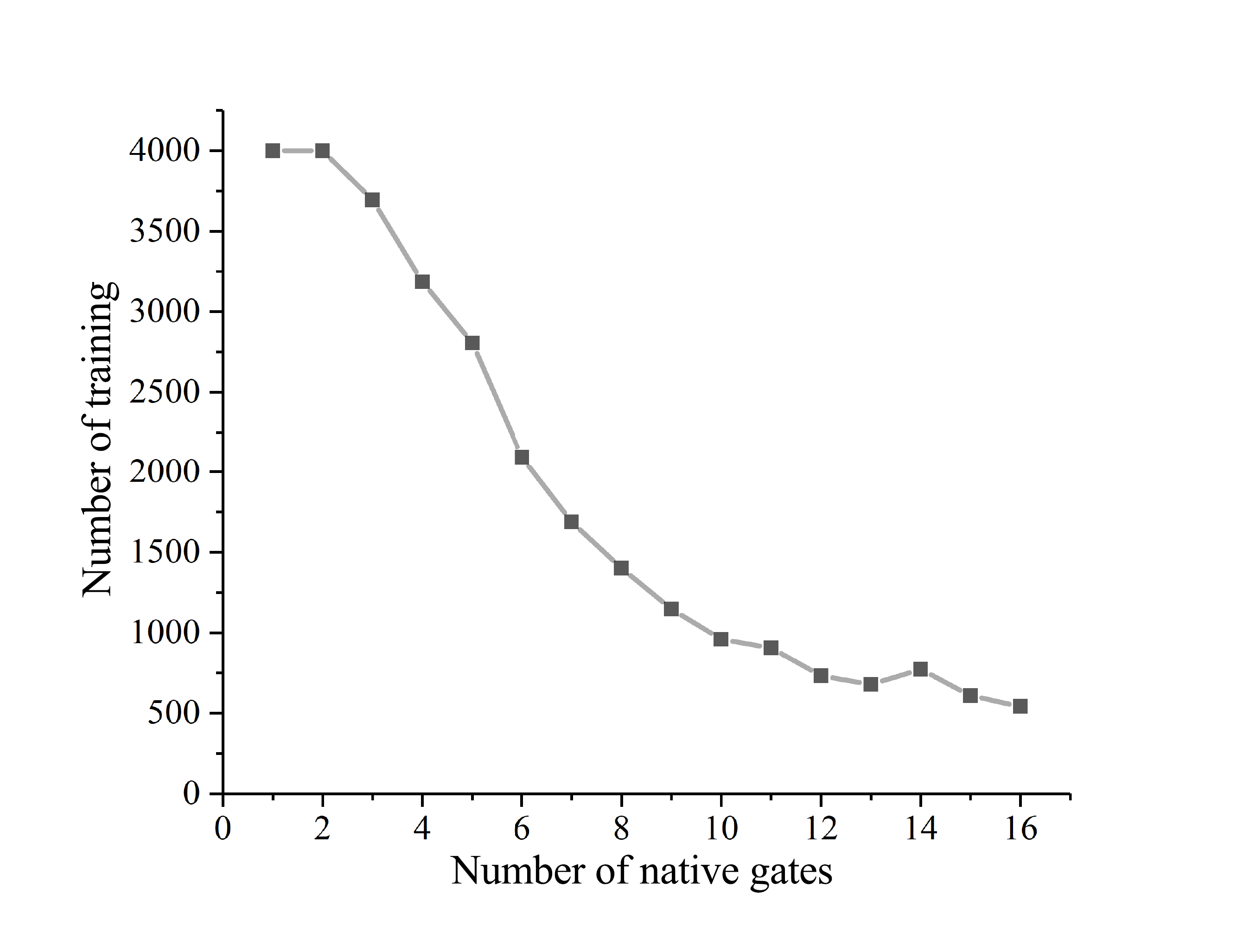}}
	\caption{\label{fig:3}The performance of the Ansatz for 32 single-qubit random reference unitaries varies with the number of employed native gates. (a) The distributions of the final errors run over the validating set with 100 random quantum states. The optimization ends when the error reaches the expected threshold (the dashed line) or the number of training rounds exceeds 4000. (b) The required average number to end the optimization as a function of the number of native gates involved in the Ansatz. The duration of the native gate is $\pi/2$ and the optimizer used here is the Adam \cite{adam} with learning rate 0.05. All parameters are ones initialized before training and are kept in the non-negative domain during the optimization to take into account experimental constraints. }
\end{figure*}

We use an Ansatz circuit to approximate the reference gate and its structure is crucial for the final result. The performance of the Ansatzes for same 32 random reference quantum gates versus the number of employed native gates is plotted in Fig.~\ref{fig:3}, whose caption describes the details, such as pulse duration and learning rate. Fig.~\ref{fig:3}(a) describes the distribution of Ansatzes' final errors compared to the associated reference quantum gates and varies with the number of native gates, where the black dashed line is a predefined error threshold, $10^{-5}$. When the validating error $\epsilon$ reaches this expected threshold (compilation succeeded) or the number of training exceeds 4000 (compilation failed), the optimization ends. 

From the evaluation results on sampled reference gates showed in Fig.~\ref{fig:3}(a), we see that once the native gate count drops below 10, there always exists points that cannot reach the error threshold, and this phenomenon comes from the limited Ansatz space due to the fact that an insufficient number of native gates cannot produce the intended reference unitary. Although two bad points (which overlap because of similar errors) are present when the native gate number is 11, they can also achieve the error requirement after 630 and 292 rounds of training
respectively by changing the learning rate from 0.05 to 0.1. When the involved native gate count reaches 12 or more, arbitrary single-qubit operations can be implemented with a fair amount of confidence. Considering more native gates would impose additional overhead for optimization and experimental implementation, we believe that an Ansatz composed of 12 native gates is a very good trade-off between approximate precision and computational overhead for arbitrary single-qubit operations.  

This argument is also supported by Fig.~\ref{fig:3}(b), which depicts the averaged number of training rounds for reaching the termination condition of optimization in terms of the applied native gate count. The required number of training rounds gradually drops off as the native gate count increases, implying that a deeper Ansatz circuit is more adequate for approximating the reference unitary.  And this trend is well maintained until the native gate count reaches 12. We emphasize that the above results are obtained under the assumption that the duration of the native gate is $\pi/2$. Other settings may deliver different results, but the final conclusion will be similar.

Quantum algorithms and error correction use a standard set of gates.  So for the implementation of algorithms, we must be able to form a universal set of such gates. For example, Hadamard transformations composed of $H$ gates are indispensable in the Grover's search algorithm, and $X$ gates are required in VQE to generate the Hartree-Fock initial states \cite{Quantum_computational_chemistry}. Table \ref{tab:A-series-of} provides a list of common single- and two-qubit standard quantum gates realized in the context of DQDs according to the MS compilation. We note, these methods can be applied in different contexts as will be demonstrated in the forthcoming subsection. After the optimization, all of the Ansatzes achieve the specified error, $10^{-5}$, compared to the associated reference gates. All parameters are initialized to 1 before training and kept in the non-negative domain during the optimization to take into account experimental constraints. The Ansatzes used for realizing single-qubit gates employ 12 single-qubit native gates with $\pi/2$ duration. The form of the Ansatz for two-qubit gate is that as depicted in Fig. \ref{fig:2}(c) in the Section \ref{sec:Method}, with pulse duration $\pi/10$ for single-qubit native gates, and $\pi/2$ for two-qubit native gates, making a final $6\pi$ runtime.  Thus this work in synergy with Ansatzes with the same runtime for an arbitrary single-qubit gate. The corresponding learning rate and number of training rounds used in optimization is collected in the second and third lines, respectively. More detailed information, e.g., the final pulse strengths, can be found in our online repository \cite{MS_compilation_gitee} in Gitee.  In general, the design of two-qubit gates is more challenging than the single-qubit operations, as revealed by the fact that the former requires a prudently selected learning rate and much longer run times for optimization. 

\begin{table}[!htbp]
	\centering
	\caption{\label{tab:A-series-of} A series of common non-parametric universal quantum gate has been realized with MS compilation in DQDs. After the number of training rounds captured in the third line, their validating errors relative to the associated reference gate all below the threshold $10^{-5}$. Every single-qubit logical gate is obtained by 12 single-qubit native gates with duration $\pi/2$. The structure of the Ansatz for two-qubit gates is depicted in Fig.~\ref{fig:2}(c) in the Section \ref{sec:Method}, with pulse duration $\pi/10$ for single-qubit native gates, and $\pi/2$ for two-qubit native gates. All parameters are initialized to 1 before training. The learning rate for optimization is captured in the second line. }
	\begin{tabular}{ccccccccc}
		\toprule 
		& H & T & S & X & Y & Z & CX & CZ\\
		\midrule
		Learning rate & 0.05 & 0.05 & 0.05 & 0.05 & 0.05 & 0.05 & 0.1 & 0.02\\
		\midrule
		Number of training & 676 & 552 & 871 & 412 & 594 & 600 & 3308 & 1162\\
		\bottomrule
	\end{tabular}
\end{table}

\begin{figure*}[!htb]
	\subfigure[ ]{\includegraphics[scale=0.52]{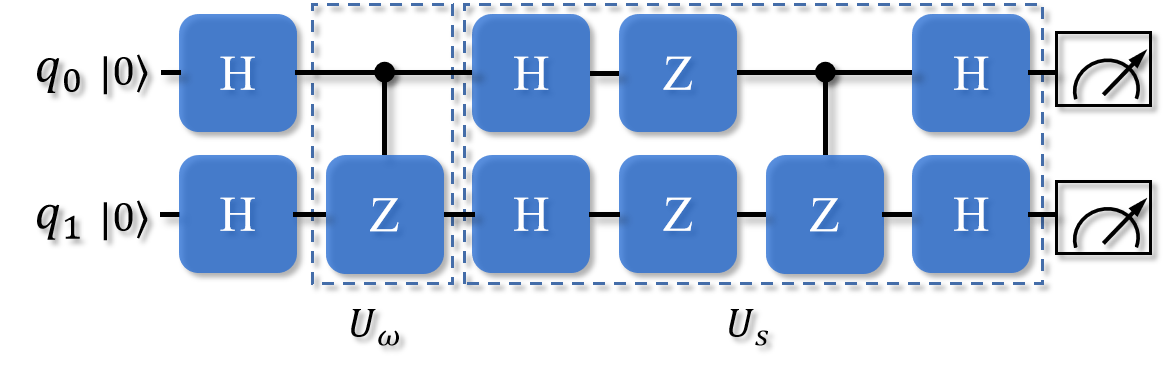}}
	\subfigure[ ]{\includegraphics[scale=0.25]{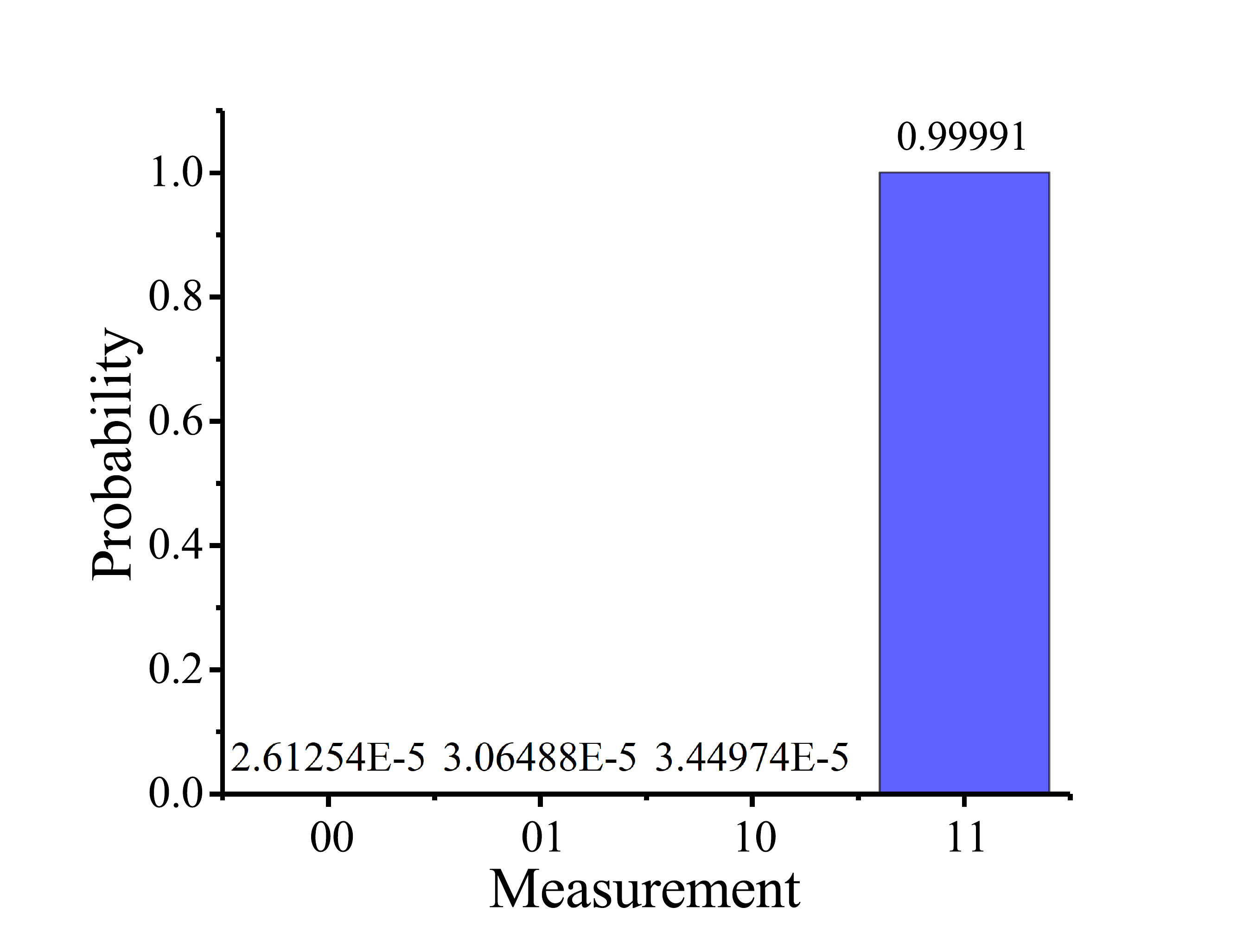}}
	\subfigure[ ]{\includegraphics[scale=0.55]{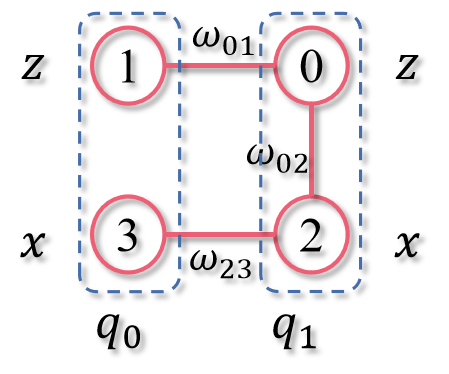}} 
	\subfigure[ ]{\includegraphics[scale=0.55]{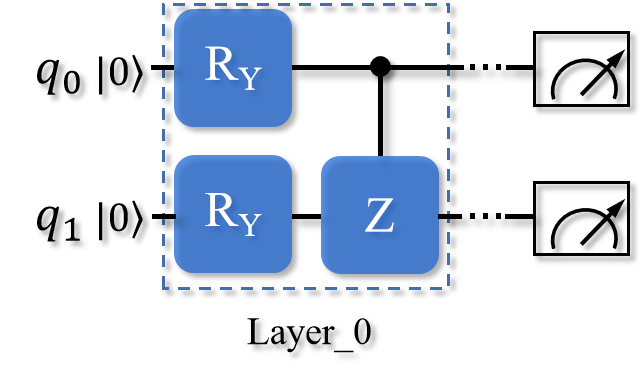}}
	\subfigure[ ]{\includegraphics[scale=0.25]{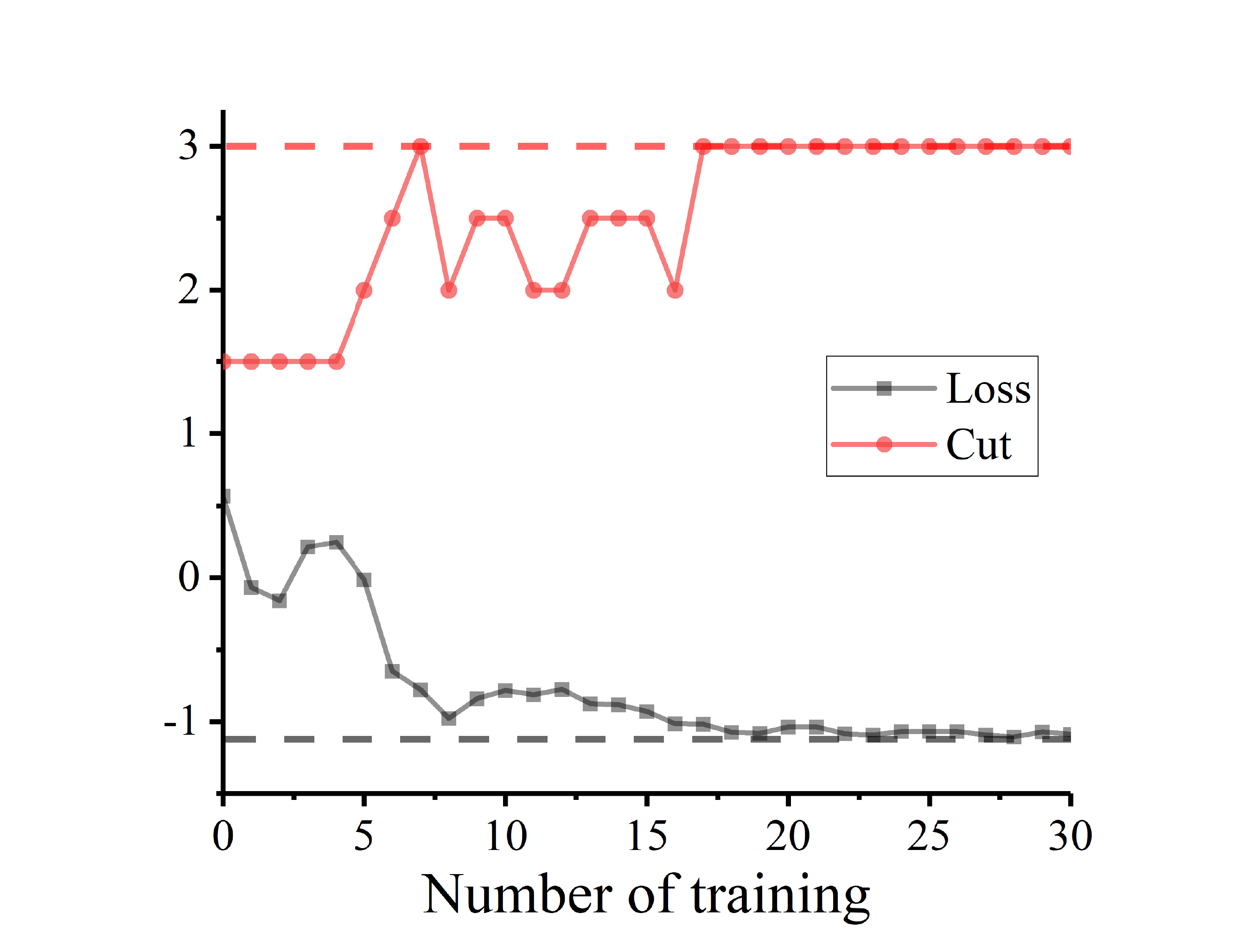}}
	\caption{\label{fig:5} Two demonstrations of MS compilation in two-qubit DQDs system: the Grover's search algorithm and the MBE-VQE for graph Max-Cut	optimization. (a) The reference circuit for Grover's search algorithm with target item ``11''. This reference circuit will be compiled, according to the MS compilation, into circuit of native gates with fixed structure and parameters before execution. (b) Numerically calculated probability distributions of different measurements, yield by the compiled version of the reference circuit (a) in DQDs system with MS compilation. (c) An undirected graph with 4 vertices and 3 edges can be represented with a two-qubit system by MBE-VQE. (d) The reference circuit with multiple basic layers used to implement a two-qubit MBE-VQE for graph Max-Cut problem. (e) The predicted cut result and loss value as functions of the number of training rounds in DQDs with MBE-VQE and MS compilation for the Max-Cut optimization of graph (c). The red and black dashed lines are the final values of cut count and loss obtained by the unrestricted system, respectively.}
\end{figure*}

\subsection{Compilation for quantum programs \label{subsec:Demonstrations-of-MS}}

In this subsection, we validate the performance of our MS compilation with two demonstrations in DQDs: the Grover's search algorithm \cite{grover_0,grover_3} and graph Max-Cut optimization \cite{max_cut}.  Our first demonstration checks whether the compiled circuit will faithfully execute the pre-designed and non-parametric reference circuit, i.e., static compilation \cite{compiler_Nature_Review}. While the second demonstration tests whether the MS compilation allows the DQDs native circuit to dynamically update its parameters efficiently to implement variational tasks, i.e., dynamic compilation \cite{compiler_Nature_Review}.

We first focus on how to implement the Grover's search algorithm in DQDs system by leveraging the MS compilation. For the problem of searching the target items from an unordered database, the quantum Grover's search algorithm permits a quadratic speedup compared to classical algorithms \cite{qc_nielsen}. Here, we consider a two-qubit Grover's search algorithm, with the target state $|\omega\rangle=|11\rangle$. The reference circuit for this task is shown in Fig.~\ref{fig:5} (a), which contains one Hadamard transformation and one Grover iteration carried out by the Oracle operator $U_{w}$ and the Flip operator $U_{s}$. All of logical gates in this reference circuit are compiled by the MS compilation into a native gate sequence which consists of only DQDs. In addition, the forms and parameters of ansatzes come directly from the results in the previous subsection and the Section \ref{sec:Method}. After execution, measurements are performed on the qubits and the probability distributions of the different results are shown in Fig.~\ref{fig:5}(b). From Fig.~\ref{fig:5}(b), we find  that the correct result ``11'' can be obtained with near unit probability.  This demonstrates that the DQDs systems with MS compilation can precisely execute pre-designed quantum programs. 

The second demonstration for our MS compilation solves the graph Max-Cut optimization in the DQDs system. The Max-Cut optimization is an NP-hard problem, which divides the vertices of an undirected graph into two parts and maximizes the sum of weighted edges being cut. For a graph with few vertices, its Max-Cut solution can be found within a reasonable amount of time by exhaustive enumeration. However, the classical algorithm will fail quickly because the resource overhead rises exponentially as the graph size scales up. Various VQA were developed to provide the quantum advantage \cite{quantum_advantage} with contemporary noisy intermediate-scale quantum devices, such as the quantum approximate optimization algorithm \cite{QAOA} (QAOA).  QAOA utilizes $n$ qubits for the Max-Cut optimization of a graph with $n$ vertices. By virtue of multi-basis graph encoding and nonlinear activation functions, based on traditional VQE \cite{VQE_photon,Quantum_computational_chemistry} and tensor networks \cite{tensor_network_0}, the multi-basis encodings VQE (MBE-VQE) improves performance using only half the quantum hardware overhead (qubit number) as well as shallower quantum circuit, and admits a quadratic reduction in measurement complexity compared to the traditional VQE \cite{MBE_for_max_cut}. 

For concreteness, we consider an undirected graph with 4 vertices and 3 edges (see Fig.~\ref{fig:5}(c) for a schematic representation). In the MBE-VQE algorithm, this graph can be represented with a two-qubit system: for example, vertices 0 and 2 are encoded into qubit $q_{0}$ and vertices 1 and 3 into qubit $q_{1}$. In addition, the vertices 0 and 1 are mapped to the $z$-axis while the vertices 2 and 3 to the $x$-axis as showed in Fig.~\ref{fig:5}(c). The loss function is made up of products of single-qubit measurements $\langle\sigma_{i}^{z}\rangle$ and $\langle\sigma_{i}^{x}\rangle$, which are dressed by function $\mathrm{tanh}(x)$ and edge-weight $\omega_{ij}$:
\begin{equation}
\begin{aligned}
Loss=&\omega_{01}\tanh(\langle\sigma_{0}^{z}\rangle)\tanh(\langle\sigma_{1}^{z}\rangle)\\&+\omega_{02}\tanh(\langle\sigma_{0}^{z}\rangle)\tanh(\langle\sigma_{0}^{x}\rangle)\\&+\omega_{23}\tanh(\langle\sigma_{0}^{x}\rangle)\tanh(\langle\sigma_{1}^{x}\rangle).
\end{aligned}
\end{equation}
The ansatz circuit in this task contains two basic layers composed of single-qubit rotation gates $R_{Y}$ and entangling two-qubit gate $CZ$, as illustrated in Fig.~\ref{fig:5}(d). Each rotation operation $R_{Y}$ is completed by 12 native gates with variable parameters and $\pi/2$ duration. While the structure and parameters for the $CZ$ gates use the results obtained in the previous subsection. According to the MBE-VQE algorithm, the predicted cut count is defined as
\begin{equation}
\begin{aligned}
\ensuremath{Cut=&\frac{\omega_{01}}{2}[1-R(\langle\sigma_{0}^{z}\rangle)R(\langle\sigma_{1}^{z}\rangle)]\\&+\frac{\omega_{02}}{2}[1-R(\langle\sigma_{0}^{z}\rangle)R(\langle\sigma_{0}^{x}\rangle)]\\&+\frac{\omega_{23}}{2}[1-R(\langle\sigma_{0}^{x}\rangle)R(\langle\sigma_{1}^{x}\rangle)],}
\end{aligned}
\end{equation}
where $R(x)$ denotes the classical rounding procedure. 

For simplicity, the weights of all the edges are taken to be $\omega_{ij}=1$. In this case, the correct solution of this Max-Cut problem is 3. Fig. \ref{fig:5}(e) displays the loss value $Loss$ and predicted cut count $Cut$ as functions of number of training rounds, where the learning rate is set be 0.1 and all native gate parameters are initially ones. The red and black dashed lines refer to the final values of $Cut$ and $Loss$ yielded by unrestricted system, respectively. From Fig.~\ref{fig:5}(e) we see that the results of the variational native circuit compiled dynamically by the MS compilation, after about 18 times of training, has almost no difference in both the predicted cut count and loss value compared to ideal results obtained in unconstrained system. This shows that, {\it by using our MS compilation, the DQD system has the potential to reliably implement variational algorithms for non-trivial applications.}

Note that these simulations were performed using MindQuantum \cite{MindQuantum}, a nascent and quickly expanding high-performance software package for quantum computation. It allows efficient problem-solving in quantum machine learning, chemistry simulation, and optimization. All detailed code and data that support this work are available in our online repository \cite{MS_compilation_gitee} for interested readers.

\section{Conclusions \label{sec:Conclusion}}

For portability, quantum algorithms are generally programmed hardware-independent, which necessitates the effective compilation of gates on platform-specific hardware for practical implementation. Due to the severe and special constraints in DQDs, what should be kept in mind during design the compilation architecture is not only the compilation of high fidelity quantum gates, but also take into account hardware constraints that, for example, prevent undesirable inter-qubit coupling.

In this work, we have shown how to use random quantum states for training in VQA for compiling or decomposing a given unitary to reduce resource overhead in loss calculation and to avoid the optimization being stuck in local optima. The performance of an ansatz consisting of various numbers of DQDs native gates to achieve single-qubit rotations was explored and the proper trade-off was found.  Using these methods, examples of common universal quantum gates were presented that showed fairly high fidelities. Most importantly, we emphasized the necessity of constant-runtime quantum gates for scalability and set up a scalable architecture, the MS compilation, which is able to layout the modularized native gates to achieve logical operations that avoid unwanted couplings between the surrounding qubits, and thus enable the faithful execution of the reference quantum circuit. Finally, to obtain an estimation of the performance, we presented two representative applications, the Grover's search algorithm for static compilation and the Max-Cut optimization for dynamic compilation. Both deliver superb results, demonstrating its feasibility for realistic implementation in the experiments. 

We believe these advances will be immensely helpful in exploiting the potential of the DQDs system to perform complex and meaningful quantum algorithms, thereby reaching a further milestone in the path towards a practical and scalable universal quantum computing device.  Furthermore, our methods could potentially be applied to a wide range of different hardware devices.  

\section*{Acknowlegment}

This work was supported by the Natural Science Foundation of Shandong Province (Grant No. ZR2021LLZ004), and the Natural Science Foundation of China (Grant No. 11475160). The author RHH would also like to thank Yang-Yang Xie, Shang-Shang Shi and Guo-Long Cui personally for fruitful discussions.  MSB was supported by the National Science Foundation of the US (MPS award No. PHYS-1820870). 

\bibliographystyle{unsrt}
\bibliography{References_library}

%
%
%
%
%
%

\end{document}